\documentclass[twocolumn,pra,letterpaper]{revtex4}
\usepackage{hyperref}
\usepackage{latexsym,amsmath,amssymb,amsfonts,graphicx,color,amsthm}
\usepackage{enumerate}
\usepackage{comment}

\newcommand{\cB}{\mathcal{B}}
\newcommand{\cC}{\mathcal{C}}

\newcommand{\cE}{\mathcal{E}}

\newcommand{\cH}{\mathcal{H}}

\newcommand{\cK}{\mathcal{K}}

\newcommand{\cM}{\mathcal{M}}
\newcommand{\cN}{\mathcal{N}}

\newcommand{\cP}{\mathcal{P}}

\newcommand{\cR}{\mathcal{R}}
\newcommand{\cS}{\mathcal{S}}
\newcommand{\cT}{\mathcal{T}}

\newcommand{\cW}{\mathcal{W}}

\newcommand{\Id}{\mathbb{I}}
\newcommand{\tr}{\text{tr}}

\newcommand{\RP}{\cR_P}

\newcommand{\Rperf}{\cR_\text{perf}}
\newcommand{\etaop}{\eta_\text{op}}
\newcommand{\Dsum}{\Delta_\text{sum}}

\newcommand{\Rop}{\cR_\text{op}}

\newcommand{\Msym}{\cM_\text{sym}}
\newcommand{\Tsym}{\cT_\text{sym}}
\newcommand{\Nsym}{\Tsym}

\newtheorem{theorem}{Theorem}
\newtheorem{lemma}[theorem]{Lemma}
\newtheorem{corollary}[theorem]{Corollary}

\newtheorem{example}{Example}

\begin{document}

\title{A simple approach to approximate quantum error correction based on the transpose channel}
\author{Hui Khoon Ng}
\altaffiliation[Current address: ]{DSO National Laboratories, and the Center for Quantum Technologies, National University of Singapore, Singapore.}
\affiliation{Institute for Quantum Information, California Institute of Technology, Pasadena, CA 91125, USA}
\author{Prabha Mandayam}
\affiliation{Institute for Quantum Information, California Institute of Technology, Pasadena, CA 91125, USA}
\date{\today}

\begin{abstract}
We demonstrate that there exists a universal, near-optimal recovery map---the transpose channel---for approximate quantum error-correcting codes, where optimality is defined using the worst-case fidelity. Using the transpose channel, we provide an alternative interpretation of the standard quantum error correction (QEC) conditions, and generalize them to a set of conditions for approximate QEC (AQEC) codes. This forms the basis of a simple algorithm for finding AQEC codes. Our analytical approach is a departure from earlier work relying on exhaustive numerical search for the optimal recovery map, with optimality defined based on entanglement fidelity. For the practically useful case of codes encoding a single qubit of information, our algorithm is particularly easy to implement.
\end{abstract}

\maketitle

\section{Introduction}\label{sec:Intro}

Quantum error correction (QEC) is one of the cornerstones of quantum information and quantum computing. Since quantum effects are extremely fragile and susceptible to damage by environmental noise, many quantum communication or computational tasks will be impossible without the use of QEC to protect the information from noise. 
QEC is thus critical for the success of quantum technologies. The idea behind QEC is a simple one---information is stored in a particular part of the system Hilbert space, cleverly chosen based on the noise process, so that a recovery operation can be applied to retrieve the information affected by the noise.

Much of the discussion in the past on error correction focuses on \emph{perfect} QEC, where the recovery operation either perfectly corrects the full CPTP noise channel, or perfectly corrects the errors conditioned on the fact that fewer than some $t$ errors occurred. However, an example of a code designed for correcting errors affected by weak amplitude damping noise presented in \cite{Leung} suggests that the requirement for perfect recovery may be too stringent for certain tasks. While the smallest known perfect QEC code requires at least five qubits to encode a single qubit, the code in \cite{Leung} uses only four qubits to achieve comparable fidelity. This illustrates a key advantage of relaxing the requirement for perfect QEC---one might be able to encode the same amount of information into fewer qubits while retaining a nearly identical level of protection from the noise process. The four-qubit code is also specially designed for the channel in question, a departure from standard QEC codes that seek to perfectly correct up to $t$ \emph{arbitrary} errors on the system. This adaptation of the code to the noise channel, an idea emphasized later in \cite{FletcherThesis}, is a crucial factor behind the success of the 4-qubit code. Such \emph{approximate} QEC (AQEC) codes reveal the possibility of designing codes that are better tailored to the particular information processing task at hand.

The analysis in \cite{Leung} was based on small perturbations of the perfect QEC conditions central to the standard theory of error correction. Subsequent work on AQEC adopted an alternate approach by recasting it as an optimization problem. One can formulate AQEC as the problem of finding the optimal encoding and recovery maps, given a noise channel and the information we want to encode (qubit or higher-dimensional object), with optimality defined in terms of a chosen measure of fidelity. In this paper, optimality is measured in terms of the \emph{worst-case fidelity}, i.e., the fidelity between the input state and the state after noise and recovery, minimized over all possible input states, for given encoding and recovery maps. This is a triple-optimization problem since one needs to optimize over all possible encodings, recovery maps and input states.

The simplest approach to solving this optimization problem is to hold either the encoding or the recovery map fixed, and then perform the optimization over the remaining two variables---the recovery or the encoding map, and the input state. The problem can be further simplified by looking instead at measures based on \emph{entanglement fidelity} \cite{Schumacher96}, and characterize the performance of the code averaged over some input ensemble. This eliminates the minimization over all input states required for the worst-case fidelity measure. The task of finding the optimal encoding or recovery map is then numerically tractable via convex-optimization methods \cite{Reimpell,Kosut06,Fletcher06, Fletcher07a,Kosut08}, but the resulting recovery is now optimal for an averaged 
measure of fidelity. Recovery maps which are near-optimal for the average entanglement fidelity have also been constructed analytically, first in \cite{BK}, and more recently in \cite{Tyson09}. 

For many communication or computational tasks, however, one would prefer an assurance that \emph{all} the information stored in the code is well-protected. In such cases, the worst-case fidelity is the appropriate measure for determining the optimality of encoding and recovery maps. The resulting double-optimization problem for a given encoding map was examined using semidefinite programming in \cite{Yamamoto}. This method however requires a relaxation of one of the constraints in the problem, so the recovery map found is typically suboptimal. Furthermore, the numerically computed recovery map is difficult to describe and understand analytically. 

In this paper, using the worst-case fidelity measure to define optimality, and assuming a fixed encoding, we construct a universal recovery map that is very easy to define analytically. This universal recovery map---the transpose channel \cite{Petzbook, BK}---gives a worst-case fidelity that cannot be too far from that of the optimal recovery. Using the fact that the transpose channel is the optimal recovery map for perfect QEC codes, we rewrite the error correction conditions \cite{EM96, BDSW96, KnillLaflamme} for perfect QEC in such a way that the role of the transpose channel is apparent. From this, we derive necessary and sufficient conditions for AQEC founded upon the transpose channel, as a natural generalization of the perfect QEC conditions. While AQEC conditions have been derived in the past from an information-theoretic perspective \cite{SW,Klesse,Buscemi,BenyTQC2,Renes10}, our conditions are algebraic, and lead to a simple and universal algorithm to find AQEC codes that does not require optimizing over all recovery maps for each encoding map. Furthermore, we demonstrate that the worst-case fidelity for the transpose channel is an easily computable quantity for the most practically useful case of codes encoding a single qubit.

Note that AQEC based on the \emph{worst-case} entanglement fidelity was also discussed recently in \cite{Beny09}, around the same time this work was done.

\section{AQEC as an optimization problem}\label{sec:Form}

Consider a physical system, with Hilbert space denoted by $\cH$. In this system, we seek to encode a qudit of information---information carried by a $d$-dimensional Hilbert space $\cH_0$, with $d\leq\text{dim}(\cH)$. In particular, we focus on the case of a \emph{subspace code}, where the qudit is encoded into a $d$-dimensional \emph{subspace} $\cC$, of $\cH$. Formally, the information is encoded into $\cC$ via a linear, invertible encoding map $\cW$. The action of noise on the system is described by a completely positive (CP), trace-preserving (TP) map $\cE:\cB(\cH)\rightarrow \cB(\cH)$. $\cE$ can describe, for example, the Markovian noise acting on the system over some time step, or the effects of a single use of a noisy channel for communication. Complete positivity of $\cE$ entails that its action can be described by a (non-unique) set of Kraus operators $\{E_i\}_{i=1}^N$, such that $\cE$ acts as $\cE(\rho)=\sum_{i=1}^N E_i\rho E_i^\dagger$. To denote the noise channel in terms of its Kraus elements, we write $\cE\sim\{E_i\}$. The fact that $\cE$ is TP is enforced by the condition $\sum_iE_i^\dagger E_i=\Id$, where $\Id$ is the identity operator for the domain of $\cE$. After the action of $\cE$, we perform a CPTP recovery map $\cR:\cB(\cH)\rightarrow\cB(\cC)$ to undo the effects of the noise, and then decode using $\cW^{-1}$.

How well the information is protected from the noise can be quantified by the \emph{fidelity} between the input qudit state and the decoded state after noise and recovery. The fidelity between any two states $\rho$ and $\sigma$ is given by $F\left(\rho,\sigma\right)\equiv \tr\sqrt{\rho^{1/2}\sigma\rho^{1/2}}$. For a pure state $\rho\equiv |\psi\rangle\langle\psi|$, this can be written as $F\left(|\psi\rangle,\sigma\right)\equiv F\left(|\psi\rangle\langle\psi|,\sigma\right)=\sqrt{\langle\psi|\sigma|\psi\rangle}$. Note that, $0\leq F(\rho,\sigma)\leq 1$, with $F=0$ if and only if $\rho$ and $\sigma$ have orthogonal support, and $F=1$ if and only if $\rho=\sigma$. The fidelity is thus a measure of how close two states are. Since we will often discuss fidelity for a state before and after the action of a map $\Phi$, we use the shorthand $F(|\psi\rangle,\Phi)\equiv F[|\psi\rangle,\Phi(|\psi\rangle\langle\psi|)]$.

Based on the fidelity measure, we say that a code $\cC$, together with $\cW$ and $\cR$, is effective at protecting the information from the noise $\cE$ if the \emph{worst-case fidelity} $\min_{\rho\in\cS(\cH_0)}F\big[\rho,\cW^{-1}\circ \cR\circ\cE \circ\cW\big]$ is close to 1. Here, $\cS(\cH_0)$ denotes the set of all states, pure or mixed, in the codespace. In fact, since the fidelity $F(\rho,\sigma)$ is jointly concave in its arguments (see, for example, \cite{NCBook}), it suffices to minimize over pure states in $\cS(\cH_0)$ only.

Above, we considered a given encoding map $\cW$ and a given recovery map $\cR$. In reality, one wants to maximize the error correction capability provided by the system by choosing $\cW$ and $\cR$ such that the worst-case fidelity is as close to 1 as possible. The problem of AQEC using a system with Hilbert space $\cH$ can thus be phrased as
\begin{equation}\label{eq:Optimize}
\max_{\cW}~\max_{\cR}\min_{|\psi\rangle\in\cH_0}F(|\psi\rangle,\cW^{-1}\circ \cR\circ\cE \circ\cW).
\end{equation}
If the quantity in Eq. \eqref{eq:Optimize} attains the maximum possible value of 1, i.e., there exist $\cW$ and $\cR$ such that the worst-case fidelity is 1, then we have perfect QEC. 

The simplest approach to solving this optimization problem is to do an exhaustive search over all possible encodings. This amounts to randomly choosing $d$-dimensional subspaces $\cC\subset\cH$. For each $\cC$, we still need to optimize over $\cR$ to maximize worst-case fidelity. For a given $\cC$, the optimization problem can be written as
\begin{equation}\label{eq:Opt}
\max_{\cR}\min_{|\psi\rangle\in\cC}F(|\psi\rangle,\cR\circ\cE),
\end{equation}
where the worst-case fidelity is computed over all pure states in $\cC$ only.

Before proceeding further, let us define some terminology. We will often make use of the square of the fidelity, which we denote as $F^2(\cdot,\cdot)\equiv \left[F(\cdot,\cdot)\right]^2$. Whenever unambiguous, we will also refer to $F^2$ as the fidelity. The recovery $\cR$ with the largest worst-case fidelity for a given $\cC$ is the \emph{optimal recovery} and is denoted by $\Rop$. The \emph{fidelity loss} $\eta_\cR$, for a given code $\cC$ and a recovery $\cR$, is defined as
\begin{equation}\label{eq:fidloss}
\eta_\cR\equiv 1-\min_{|\psi\rangle\in\cC}F^2\left[|\psi\rangle,(\cR\circ\cE)(|\psi\rangle\langle\psi|)\right].
\end{equation}
The fidelity loss for $\Rop$, denoted $\etaop$, is $\etaop=\min_\cR\eta_\cR$. We refer to $\etaop$ as the \emph{optimal fidelity loss}. A code $\cC$ is said to be $\epsilon$-\emph{correctable} if it has $\etaop\leq \epsilon$ for some $\epsilon\in[0,1]$. $\epsilon$-correctable codes with $\epsilon\ll 1$ are approximately correctable in the sense that code states have fidelity at least $\sqrt{1-\epsilon}\simeq 1-\epsilon/2$ after the action of the noise and (optimal) recovery.

\section{Transpose channel as universal, near-optimal recovery}\label{sec:TransposeChannel}
Here, we describe the transpose channel and demonstrate that it is indeed the standard recovery map for perfect QEC codes characterized by the well-known QEC conditions. We then proceed to show that the transpose channel is nearly optimal even in the case of AQEC codes.

\subsection{Transpose channel}

Consider a $d$-dimensional code $\cC$, and a CPTP noise channel $\cE\sim\{E_i\}_{i=1}^N$. Let $P$ be the projector onto $\cC$ and $P_\cE$ be the projector onto $\cP_\cE\equiv$ the support of $\cE(\cC)$. The transpose channel is the CPTP map $\RP:\cB(\cP_\cE)\rightarrow \cB(\cC)$ such that
\begin{equation}\label{eq:transpose}
\RP\left(\cdot\right)\equiv \sum_{i=1}^N PE_i^\dagger \cE(P)^{-1/2}\left(\cdot\right)\cE(P)^{-1/2}E_iP,
\end{equation}
where the inverse of $\cE(P)$ is taken on its support. The transpose channel can be understood as being composed of three CP maps: $\RP=\cP\circ\cE^\dagger\circ\cN$, where $\cP$ is the projection $P(\cdot)P$ onto $\cC$, $\cE^\dagger$ is the adjoint of $\cE$, and $\cN$ is the normalization map $\cN(\cdot)=\cE(P)^{-1/2}(\cdot)\cE(P)^{-1/2}$. In this form, $\RP$ is manifestly independent of the choice of Kraus representation for $\cE$.

$\RP$ is a special case of a recovery map introduced in \cite{BK} for reversing the effects of a quantum channel on a given initial state. $\RP$ defined here is exactly the case for the initial state $P/d$, where $d$ is the dimension of $\cC$. In \cite{IPSPaper}, $\RP$ was shown to be useful for correcting information carried by codes preserved according to an operationally motivated notion. The term \textit{transpose channel} owes its origin to \cite{Petzbook}, where this channel was first defined in an information-theoretic context. It was shown \cite{Petz03} that the transpose channel has the property of being the unique noise channel that saturates Uhlmann's theorem i.e. the monotonicity of relative entropy---a fact that was later used to characterize states that saturate the strong subbadditivity of quantum entropy \cite{HJPW04}.

While our focus is on AQEC, understanding the relevance of $\RP$ to perfect QEC provides the intuition behind the AQEC conditions presented later. An important characterization of perfect QEC codes is the set of QEC conditions \cite{EM96, BDSW96, KnillLaflamme}, which we briefly review here (see, for example, \cite{NCBook}):
\begin{theorem}[{\bf Perfect QEC conditions}]\label{thm:ECcond}
A CPTP recovery $\cR$ that perfectly corrects a CP map $\cE$ on a subspace code $\cC$ exists if and only if
\begin{equation}\label{eq:ECcond}
\forall i,j,\quad PE_i^\dagger E_jP=\alpha_{ij}P,
\end{equation}
for some complex matrix $\alpha$.
\end{theorem}
It is useful to rewrite Eq. \eqref{eq:ECcond} in a ``diagonal" form. $\alpha$ is clearly Hermitian, and can be diagonalized with a unitary $u$ such that $\alpha=uD u^\dagger$, where $D$ is the diagonal matrix of eigenvalues. The set $\{F_k\equiv \sum_iu_{ik}E_i\}$ constitutes a different Kraus representation for $\cE$. With this choice of Kraus representation, the perfect QEC conditions take the diagonal form
\begin{equation}\label{eq:diagQEC}
\forall k,l,\quad PF_k^\dagger F_lP=\delta_{kl}d_{kk}P,
\end{equation}
where $d_{kk}$ are the diagonal entries of $D$.

The recovery map $\cR$ when Eq. \eqref{eq:ECcond} is satisfied---which we denote as $\Rperf$---is constructed as follows \cite{NCBook}: using the polar decomposition $F_kP=\sqrt{d_{kk}}U_kP$, $\Rperf:\cB(\cP_\cE)\rightarrow \cB(\cC)$ is given by $\Rperf\sim\{PU_k^\dagger\}$. One can check that $\Rperf$ is TP on its domain $\cB(\cP_\cE)$, and that for any $\rho\in \cB(\cC)$, $(\Rperf\circ\cE)(\rho)=\Big(\sum_kd_{kk}\Big)\rho$. From the QEC conditions (Eq. \eqref{eq:diagQEC}), we see that $\sum_kd_{kk}=\tr[\cE(\rho)]$ is independent of $\rho$, and is exactly equal to 1 if and only if $\cE$ is TP on $\cC$. $\Rperf$ thus recovers the original code state, up to any reduction in trace due to the possible non-TP nature of $\cE$.

A natural question to ask here is how the transpose channel $\RP$ relates to the recovery $\Rperf$ for a given $\cE$ and $\cC$ that satisfy the QEC conditions. Here, we show that they are exactly the same map, as previously noted in \cite{BK}:
\begin{lemma}\label{lem:BKrec}
$\RP=\Rperf$.
\end{lemma}
\begin{proof}
Observe that $\cE(P)=\sum_k(F_kP)(PF_k^\dagger)=\sum_kd_{kk}P_k$, where $P_k\equiv U_kPU_k^\dagger$. Eq. \eqref{eq:diagQEC} gives $PU_k^\dagger U_lP=\delta_{kl}P$, so that $P_k$'s are orthogonal projectors with $P_kP_l=\delta_{kl}P_k$. Hence, $\cE(P)^{-1/2}=\sum_kP_k/\sqrt{d_{kk}}$. The Kraus operators $\{PF_k^\dagger \cE(P)^{-1/2}\}$ of $\RP$ can hence be written as
\begin{align}\label{eq:cEPKraus}
PF_k^\dagger \cE(P)^{-1/2}&=\sum_l\sqrt{d_{kk}/d_{ll}}PU_k^\dagger U_lPU_l^\dagger=PU_k^\dagger,
\end{align}
which are exactly the Kraus operators of $\Rperf$.
\end{proof}

Perfect QEC is often discussed for a noise channel that is CP but not necessarily TP. In fact, Theorem \ref{thm:ECcond} and Lemma \ref{lem:BKrec} remain true even for a non-TP $\cE$. The non-TP scenario is particularly relevant when we deal with a system of $n$ quantum registers, where each register is independently affected by some CPTP noise $\cE_1$. One often looks for codes that perfectly correct the noise up to some maximum number $t$ of quantum registers with errors. Then, instead of having $\cE\equiv \cE_1^{\otimes n}$, the relevant noise channel for perfect QEC describes noise where at most $t$ registers have errors. Such an $\cE$ is not TP, since we have discarded the part of $\cE_1^{\otimes n}$ that corresponds to having errors in more than $t$ registers.

Actually, a perfect QEC code for such a non-TP noise channel can be viewed as an AQEC code for the original $n$-register noise channel $\cE_1^{\otimes n}$, which \emph{is} TP. In our AQEC discussion, the code we look for is approximately correctable on the channel anyway, so $\cE$ is always assumed to be TP, which is often the physically-relevant scenario. The TP requirement is also important for fidelity to be a good measure of the efficacy of the recovery operation. Note that the analysis in the remainder of the paper applies to a special type of non-TP maps---$\cE\sim\{E_i\}$ such that $\sum_i P E_i^\dagger E_iP=aP$ for $0\leq a \leq 1$, giving an additional proportionality factor $a$ in our expressions.

\subsection{Near-optimality of the transpose channel}
In general, $\RP$ need not be the optimal recovery map $\Rop$ for a given $\cC$ and $\cE$. However, in the following theorem  and the subsequent corollary, which form the core results of our paper, we show that it does not do much worse than $\Rop$.
\begin{theorem}\label{thm:Fid}
Consider a $d$-dimensional code $\cC$ with optimal fidelity loss $\etaop$ under a CPTP noise channel $\cE$. For any $|\psi\rangle\in\cC$,
\begin{align}\label{eq:thmFid}
F^2(|\psi\rangle,\Rop\circ\cE)\leq \sqrt{1+(d-1)\etaop}F(|\psi\rangle,\RP\circ\cE).
\end{align}
\end{theorem}
\noindent{\it Proof.} Let $\{R_j\}$ be a set of Kraus operators of $\Rop:\cB(\cP_\cE)\rightarrow\cB(\cC)$. For any $|\psi\rangle\in\cC$, following \cite{BK}, we have
\begin{align}
&\qquad F^2(|\psi\rangle,\Rop\circ\cE)\label{eq:fid2}\\
&\leq \sqrt{\big(\sum_i|\langle E_i^\dagger \cE(P)^{-1/2}E_i\rangle|^2\big)\big(\sum_j|\langle R_j\cE(P)^{1/2}R_j^\dagger\rangle|^2\big)},\notag
\end{align}
where $\langle\cdot\rangle$ denotes expectation value with respect to $|\psi\rangle$. Since $\Rop$ is TP, we have that $\sum_j|\langle R_j\cE(P)^{1/2}R_j^\dagger\rangle|^2\leq \langle\sum_jR_j\cE(P)R_j^\dagger\rangle = \langle(\Rop\circ\cE)(P)\rangle$.

Now, choose a basis $\{|\psi_i\rangle\}_{i=1}^d$ for $\cC$ with $|\psi_1\rangle\equiv|\psi\rangle$. Let $\rho_i\equiv(\Rop\circ\cE)(|\psi_i\rangle\langle\psi_i|)=\sum_{kl}\alpha_{kl}^{(i)}|\psi_k\rangle\langle\psi_l|$, for coefficients satisfying $\sum_k\alpha_{kk}^{(i)}=1$ and $\alpha_{kk}^{(i)}\geq 0~\forall k$. From the definition of $\etaop$, $\alpha_{ii}^{(i)}=\langle\psi_i|\rho_i|\psi_i\rangle\geq 1-\etaop$. This implies $\sum_{k\neq i}\alpha_{kk}^{(i)}\leq \etaop$, which in turn gives $\alpha_{kk}^{(i)}\leq \etaop~\forall k\neq i$. Since $|\psi\rangle=|\psi_1\rangle$ by construction, we get $\sum_j|\langle R_j\cE(P)^{1/2}R_j^\dagger\rangle|^2\leq \langle(\Rop\circ\cE)(P)\rangle\leq 1+(d-1)\etaop$. Putting this into Eq. \eqref{eq:fid2}, and noting that $\big(\sum_i|\langle E_i^\dagger \cE(P)^{-1/2}E_i\rangle|^2\big)^{1/2}\leq F(|\psi\rangle,\RP\circ\cE)$, gives Eq. \eqref{eq:thmFid}.\qed

Let $\eta_P$ be the fidelity loss for code $\cC$ with $\RP$ as the recovery. Then, Theorem \ref{thm:Fid} implies
\begin{corollary}\label{cor:iff}
$\eta_P$ satisfies $\etaop\leq \eta_P\leq \etaop f(\etaop;d)$, where $f(\eta;d)$ is the function
\begin{equation}\label{eq:funcf}
f(\eta;d)\equiv \frac{(d+1)-\eta}{1+(d-1)\eta}=(d+1)+O(\eta).
\end{equation}
\end{corollary}
\begin{proof}
$\eta_P\geq \etaop$ is true by definition of $\etaop$. For any $|\psi\rangle\in\cC$, let $F^2(|\psi\rangle,\RP\circ\cE)\equiv 1-\eta_{P,\psi}$. Then, by definition, the fidelity-loss is $\eta_P=\max_\psi(\eta_{P,\psi})$. From Theorem \ref{thm:Fid}, $1-\etaop\leq F^2(|\psi\rangle,\cR_\text{op}\circ\cE)\leq \sqrt{[1+(d-1)\etaop](1-\eta_{P,\psi})}$.
Rearranging gives $\eta_{P,\psi}\leq \etaop f(\etaop;d)$. Since this holds for all $\eta_{P,\psi}$, it also holds for $\eta_P$.
\end{proof}
The inequality $\eta_P\leq \etaop f(\etaop;d)$ makes precise our statement that $\RP$ is near-optimal, with the additional factor of $~(d+1)$. For the most practically-relevant case of a code encoding a single qubit, this is only a factor of 3. Note that, for $\etaop=0$, the inequality in Corollary \ref{cor:iff} collapses to $\eta_P=\etaop$, as expected from Lemma \ref{lem:BKrec}. Corollary \ref{cor:iff} provides necessary and sufficient conditions for $\cC$ to be approximately correctable---$\cC$ is approximately correctable if and only if $\eta_P$ is small.

We do not know if the upper bound on $\eta_P$ in Corollary \ref{cor:iff} is tight. However, the appearance of the dimension $d$ of the code in the bound is unavoidable, as can be seen from the following example:
\begin{example}
Consider a noise channel $\cE$, whose action on a code $\cC$ is given by the set of Kraus operators $\{E_iP\}=\{\sqrt{1-p}~P,\sqrt p ~|0\rangle\langle 0|,\sqrt p ~|0\rangle\langle 1|,\ldots, \sqrt p ~|0\rangle\langle d-1|\}$, for $0\leq p\ll 1$. $\cE$ acts like the identity channel on $\cC$, except for a small damaging component that maps a small part of every code state onto $|0\rangle$. For $d\geq 3$, one can show that the worst-case fidelity, when using $\RP$ as the recovery, occurs for state $|0\rangle$. The corresponding fidelity loss is $\eta_P=(d-1)p/[1+(d-1)p]$. Since $\cE$ is nearly the identity channel, we might instead do nothing (identity channel as the recovery), for which the fidelity loss is $\eta_0=p$. $\eta_0$ is always smaller than $\eta_P$ for small $p$. Since $\etaop\leq \eta_0$, we see that $\eta_P/\etaop\geq \eta_P/\eta_0=(d-1)/[1+(d-1)p]$, which grows as $d$ increases, for fixed $p$. Hence, for this noise channel and code, the separation between $\eta_P$ and $\etaop$ grows as $d$ increases.
\end{example}

That the dimension of the code space appears here is perhaps not surprising. In the next section, we will see that this approach to AQEC using the transpose channel can be viewed as a perturbation from the perfect QEC case. The factor of $d$ appearing in our bounds can hence be understood as quantifying the number of degrees of freedom in which the approximate case can deviate from the perfect case. 

Note, however, that as $d$ gets large, $f(\eta;d)$ approaches $1/eta$. In this case, the inequality in Corollary 4 simply becomes the trivial statement $\eta_op\leq \eta_P\leq 1$. While we will often only be interested in
codes with small values of $d$, this demonstrates the weakness in the bounds derived here for large $d$ values.

\section{The transpose channel and QEC conditions}\label{sec:Conds}

One of the key tools in perfect QEC are the perfect QEC conditions (Theorem \ref{thm:ECcond}). Conditions characterizing AQEC codes would likewise be useful. A natural approach is to perturb the perfect QEC conditions to allow for small deviations. For example, the four-qubit code for the amplitude damping channel in \cite{Leung} was shown to obey a set of perturbed QEC conditions. More recently, \cite{MD} examined small perturbations of the perfect QEC conditions for general CPTP channels. However, the analysis in \cite{MD} is often complicated, and one wonders if there is a simpler approach using the transpose channel. In this section, we discuss such a set of AQEC conditions built upon Corollary \ref{cor:iff}. We begin by first writing down an alternate, but equivalent set of perfect QEC conditions which highlights the role of the transpose channel:
\begin{theorem}[{\bf Alternate form of perfect QEC conditions}]\label{thm:ECcond2}
A code $\cC$ satisfies the perfect QEC conditions (Theorem \ref{thm:ECcond}) if and only if it satisfies
\begin{equation}\label{eq:ECcond2}
\forall i,j,\quad PE_i^\dagger \cE(P)^{-1/2}E_j P=\beta_{ij}P,
\end{equation}
where $\beta\equiv \sqrt\alpha$, for $\alpha$ is defined in Eq.\eqref{eq:ECcond}.
\end{theorem}
\begin{proof}
For a code $\cC$ that satisfies the perfect QEC conditions (Theorem \ref{thm:ECcond}), using Eq. \eqref{eq:cEPKraus} and $PU_k^\dagger U_lP=\delta_{kl}P$, we have
\begin{equation}\label{eq:ECcond3}
PF_k^\dagger\cE(P)^{-1/2}F_lP
=\delta_{kl}\sqrt{d_{kk}}P.
\end{equation}
This diagonal form can be rotated to any other Kraus representation using a unitary $u$ so that $F_k=\sum_iu_{ik}E_i$ and $\alpha=uDu^\dagger$. Defining $\beta\equiv \sqrt\alpha$ gives Eq. \eqref{eq:ECcond2}, thus showing that if a code $\cC$ satisfies the perfect QEC conditions, it also satisfies Eq. \eqref{eq:ECcond2}.

Conversely, suppose we start from the diagonal form of Eq. \eqref{eq:ECcond2} as in Eq. \eqref{eq:ECcond3}, which can be accomplished by choosing $u$ so that $\beta$ is diagonal with entries $\sqrt{d_{kk}}$. Then taking the square root of Eq. \eqref{eq:ECcond3} gives $\cE(P)^{-1/4}F_kP=\left(d_{kk}\right)^{1/4}V_kP$ for some unitary $V_k$, so that $F_kP=\left(d_{kk}\right)^{1/4}\cE(P)^{1/4}V_kP$. Putting this into Eq. \eqref{eq:ECcond3} gives $PV_k^\dagger V_lP=\delta_{kl}P$. Furthermore, $\cE(P)^{1/2}=[\sum_k(F_kP)(PF_k^\dagger)]^{1/2}=\sum_k\sqrt{d_{kk}}V_kPV_k^\dagger$. Direct computation then gives $PF_k^\dagger F_lP=\delta_{kl}d_{kk}P$, which is exactly Eq. \eqref{eq:diagQEC}. Applying an appropriate $u$ to rotate to the desired Kraus representation gives Eq. \eqref{eq:ECcond}.
\end{proof}

Observe that the left-hand side of Eq. \eqref{eq:ECcond2} is a Kraus operator of $\RP\circ\cE$. Thus, the QEC conditions in Theorem \ref{thm:ECcond2}, and equivalently the original conditions stated in Theorem \ref{thm:ECcond}, simply express the fact that $\cC$ is perfectly correctable if and only if $\RP\circ\cE\propto\hat{\cP}$. The 
proportionality factor is $\sum_{ij}\beta_{ij}^2=\sum_{ij}\alpha_{ij}=\sum_kd_{kk}$.

We can now obtain a set of conditions for AQEC by perturbing this alternate form of the QEC conditions.
\begin{theorem}[{\bf AQEC conditions}]\label{thm:AECcond}
Consider a CPTP channel $\cE\sim\{E_i\}$, and a $d$-dimensional code $\cC$ with projector $P$. Let $\Delta_{ij}\in\cB(\cC)$ be traceless operators such that
\begin{equation}\label{eq:AECcond}
PE_i^\dagger \cE(P)^{-1/2}E_j P=\beta_{ij}P+\Delta_{ij},
\end{equation}
where $\beta_{ij}\in\mathbb{C}$. Then, for $\epsilon\in[0,1]$, $\exists \; \eta\in[0,1]$ given by
\begin{equation}\label{eq:etaP}
\eta =\max_{|\psi\rangle\in\cC}\sum_{ij}\left[\langle\psi|\Delta_{ij}^\dagger\Delta_{ij}|\psi\rangle-\left\vert\langle\psi|\Delta_{ij}|\psi\rangle\right\vert^2\right].
\end{equation}
such that
\begin{itemize}
\item[(i)] $\cC$ is $\epsilon$-correctable \emph{if} $\eta\leq \epsilon$;
\item[(ii)] $\cC$ is $\epsilon$-correctable \emph{only if} $\eta\leq \epsilon f(\epsilon;d)$, where $f$ is the function defined in Eq. \eqref{eq:funcf}.
\end{itemize}
\end{theorem}
\begin{proof} 
The left-hand side of Eq. \eqref{eq:AECcond} is a Kraus operator of $\RP\circ\cE$. This, along with the TP condition for $\RP\circ\cE$, gives the expression on the right-hand side of Eq. \eqref{eq:etaP} for $\eta_P$. Setting $\eta=\eta_P$, conditions (i) and (ii) follow from Corollary \ref{cor:iff}.
\end{proof}
\noindent Eq. \eqref{eq:etaP} elucidates how the fidelity loss arises from the presence of the $\Delta_{ij}$ operators. If $\Delta_{ij}=0~\forall i,j$, we have perfect QEC.

The AQEC conditions, like the perfect QEC conditions, provide a way to check if a code is approximately correctable, without requiring knowledge of the optimal recovery. More precisely, given a maximum tolerable fidelity loss $\epsilon$ for some information processing task at hand, one can check if a code $\cC$ is $\epsilon$-correctable as follows. The AQEC conditions instruct us to compute $\eta_P$, which can be done once we know $\cC$ and the noise channel $\cE$. If $\eta_P\leq \epsilon$, then $\cC$ is a good code. If however, $\eta_P$ violates the inequality in condition (ii), we know that $\cC$ is not good enough for our purposes. Of course, there is a gap---for $\eta_P$ taking values $\epsilon\leq \eta_P\leq\epsilon f(\epsilon;d)$, we cannot use the conditions to determine if $\cC$ is within our tolerable fidelity loss, but this gap is small for small $d$. We do not know if the gap can be shrunk by replacing $\eta_P$ with the fidelity loss for a different recovery map than the transpose channel, but we believe it is unlikely to vanish completely.

For a general $\cC$, the fidelity loss $\eta_P$ may be difficult to compute as it requires a maximization over all states in the code space. However, there is a quick way to check for sufficiency by relaxing condition (i) of Theorem \ref{thm:AECcond}:
\begin{corollary}
$\cC$ is $\epsilon$-correctable for some $\epsilon\in[0,1]$ if $\Vert\Dsum~\Vert\leq \epsilon$, where $\Dsum\equiv\sum_{ij}\Delta_{ij}^\dagger \Delta_{ij}$, and $\Vert\cdot\Vert$ denotes the operator norm.
\end{corollary}
\begin{proof}
Observe that the right-hand side of Eq. \eqref{eq:etaP} satisfies $\sum_{ij}[\langle\psi|\Delta_{ij}^\dagger\Delta_{ij}|\psi\rangle-\vert\langle\psi|\Delta_{ij}^\dagger|\psi\rangle\vert^2]\leq \langle \psi|\Dsum|\psi\rangle$. Maximizing this expression over all $|\psi\rangle\in\cC$ gives $\Vert\Dsum\Vert$. Hence, $\eta_P\leq \Vert\Dsum\Vert$, and the sufficiency condition (i) in Theorem \ref{thm:AECcond} is satisfied if $\Vert\Dsum\Vert\leq \epsilon$.
\end{proof}
\noindent Since $\Dsum\geq 0$, its operator norm is given by its maximum eigenvalue, which is easily computable. In fact, for codes encoding a single qubit, it is easy to show (using the Pauli basis, for example) that $\Vert\Dsum\Vert=1-\sum_{ij}\vert\beta_{ij}\vert^2$. Note that $\beta_{ij}$ for any code $\cC$ and noise channel $\cE$ is simply given by $\beta_{ij}=(1/d)~\tr(PE_i^\dagger \cE(P)^{-1/2}E_jP)$.

\section{Computing $\eta_P$ for qubit codes}\label{sec:Qubit}

Computing $\eta_P$ for a general code requires an exhaustive optimization over all states in the code. However, for the practically relevant case of codes encoding a single qubit, i.e., $\cC$ with dimension $d=2$, $\eta_P$ turns out to require only simple eigen-analysis to compute. 

For a qubit code, $(\RP\circ\cE):\cB(\cC)\rightarrow\cB(\cC)$ is a qubit map. Observe that $\RP\circ\cE$ is not only CPTP but also unital (i.e., $(\RP\circ\cE)(P)=P$). Here, we show that the worst-case fidelity for a unital, CPTP qubit map is easy to compute. While our context requires only a unital, CPTP qubit map, we begin with a general CP map $\Phi\sim\{K_i\}$ on a $d$-dimensional Hilbert subspace $\cC$, so as to highlight why the qubit case is particularly simple. 

We begin by choosing a Hermitian basis  $\{O_0,O_1,\ldots, O_{d^2-1}\}$ for $\cB(\cC)$ where $O_0\equiv \Id,~ O_\alpha^\dagger =O_\alpha~\forall \alpha,~\tr\{O_\alpha^\dagger O_\beta\}=\delta_{\alpha\beta} d~~ \forall \; \alpha,\beta$. The operators $\{O_\alpha, \alpha=1,\ldots,d^2-1\}$ are clearly traceless. Such a basis exists for any $d$---for example, one can use the set of standard generators of the $SU(d)$ group, augmented with the identity operator. The action of $\Phi$ can be represented as a matrix $\cM$, acting on vectors (operators in $\cB(\cC)$) in the Hilbert-Schmidt space, with matrix elements
\begin{equation}\label{eq:MPhigen}
\cM_{\alpha\beta}\equiv \frac{1}{d}\tr\{O_\alpha\Phi(O_\beta)\}.
\end{equation}
Since $\Phi$ is CP and $O_\alpha$'s are Hermitian, $\cM_{\alpha\beta}^*=\cM_{\alpha\beta}$, so $\cM$ is a real matrix. 

Now, the density operator corresponding to any pure state $|\psi\rangle$ in $\cC$ can be expanded in terms of the Hermitian basis as
\begin{equation}\label{eq:Bloch}
|\psi\rangle\langle\psi|=\frac{1}{d}\left(\Id+\mathbf{s}\cdot\mathbf{O}\right)=\frac{1}{d}~\vec s\cdot\vec O,
\end{equation}
where $\mathbf{s}$ is a real $(d^2-1)$-element vector, $\vec s\equiv (1,\mathbf{s})$, $\mathbf{O}\equiv (O_1,O_2,\ldots, O_{d^2-1})$, and $\vec O\equiv (\Id,\mathbf{O})$. $\mathbf{s}$ is not an arbitrary vector, but in general has to obey some constraints in order for it to correspond to a pure state.

Using Eqs. \eqref{eq:MPhigen} and \eqref{eq:Bloch}, the fidelity for a state $|\psi\rangle\in\cC$ under the map $\Phi$ can be written as $F^2\big[|\psi\rangle,\Phi(|\psi\rangle\langle\psi|)\big]=\frac{1}{d} ~s^T\cM~ s$, where $s$ is $\vec s$ written as a column vector, and the superscript $T$ denotes the transpose. We can rewrite the expression for the fidelity using the symmetrized version of $\cM$: $\Msym\equiv \frac{1}{2}(\cM+\cM^T)$. Observe that $s^T\Msym~s=s^T\cM~s$. Therefore,
\begin{equation}\label{eq:MFidgen1}
F^2\big(|\psi\rangle,\Phi(|\psi\rangle\langle\psi|\big)=s^T\Msym~s.
\end{equation}
Finding the worst-case fidelity is hence equivalent to the following minimization problem for a real, symmetric matrix $\Msym$:
\begin{subequations}\label{eq:Fmin}
\begin{align}
\label{eq:Fidgen}\text{minimize: }&\quad s^T \Msym ~s,\\
\label{eq:constrgen}\text{constraint: }&\quad s \text{ corresponds to a pure state}.
\end{align}
\end{subequations}
For $d>2$, the constraint Eq. \eqref{eq:constrgen} is difficult to write down. Even if we relax the constraint to include mixed states, it is not known in general what $s$ corresponding to a (positive, trace-1) density operator looks like. This constrained minimization problem is hence not simple for a general $d$.

For qubits ($d=2$) however, the constraint equation \emph{is} simple to write down. We choose the operator basis to be the Pauli basis. Given an orthonormal basis $\{|v_1\rangle,|v_2\rangle\}$ for the qubit code space, the Pauli basis $\{\sigma_0\equiv\Id_2,\sigma_x,\sigma_y,\sigma_z\}$ can be constructed as
\begin{align}\label{eq:Pauli}
\sigma_0&=|v_1\rangle\langle v_1|+|v_2\rangle\langle v_2|\equiv\Id_2,\notag\\
\sigma_x &=|v_1\rangle\langle v_2|+|v_2\rangle\langle v_1|,\notag\\
\sigma_y &=-i(|v_1\rangle\langle v_2|-|v_2\rangle\langle v_1|),\notag\\
\text{and}\quad\sigma_z &=|v_1\rangle\langle v_1|-|v_2\rangle\langle v_2|.
\end{align}
Eq. \eqref{eq:Bloch} then corresponds to the Bloch sphere representation of a pure state, with the Bloch vector $\mathbf{s}\equiv(s_x,s_y,s_z)$ satisfying $\Vert\mathbf{s}\Vert=(s_x^2+s_y^2+s_z^2)^{1/2}=1$. The constraint Eq. \eqref{eq:constrgen} becomes
\begin{equation}\tag{\ref{eq:constrgen}$'$}
\text{constraint: }\quad s=(1,\mathbf{s}),\quad\text{ with }\Vert\mathbf{s}\Vert=1.
\end{equation}
The constrained minimization problem can then be solved using the Lagrange multiplier method.

For the case of a CPTP qubit map that is also unital, the minimization problem can be further simplified. For any CPTP, unital $\Phi$ (arbitrary $d$), $\cM$ takes the form
\begin{equation}\label{eq:MPhi}
\cM=\left(
\begin{array}{c|ccc}
1&0&\ldots&0\\
\hline
0&&&\\
\vdots&&\cT&\\
0&&&
\end{array}
\right).
\end{equation}
The first row comes from the fact that $\Phi$ is TP, and the first column from the fact that $\Phi$ is unital. $\cT$ is a $(d-1)\times (d-1)$ real matrix. Defining $\Nsym\equiv \frac{1}{2}(\cT+\cT^T)$, Eq. \eqref{eq:MFidgen1} can be written as $F^2\big(|\psi\rangle,\Phi(|\psi\rangle\langle\psi|\big)=\frac{1}{d}(1+\mathbf{s}^T\Nsym~\mathbf{s})$. This means that we can equivalently minimize $\mathbf{s}^T\Nsym~\mathbf{s}$ instead of the original $s^T\Msym~s$ in Eq. \eqref{eq:Fidgen}. For a qubit CPTP, unital $\Phi$ then, the constrained minimization problem becomes
\begin{subequations}\label{eq:Fminqb}
\begin{align}
\label{eq:Fidqb}\text{minimize: }&\quad \mathbf{s}^T \Nsym ~\mathbf{s},\\
\label{eq:constrqb}\text{constraint: }&\quad \Vert\mathbf{s}\Vert=\sqrt{s_x^2+s_y^2+s_z^2}=1.
\end{align}
\end{subequations}
This simply tells us to minimize the expectation value of $\Nsym$ with respect to all real \emph{unit} vectors $\mathbf{s}$.

Since $\Nsym$ is real and symmetric, it can be diagonalized with an orthogonal matrix $Q$ so that $\Nsym=Q^T\cT_D Q$, where $\cT_D$ is a real, diagonal matrix of eigenvalues of $\Nsym$. Then $s^T\Nsym ~s=(Qs)^T\cT_D (Qs)$. $Q$, being orthogonal, preserves the length of the vector it acts on. The minimization problem Eq. \eqref{eq:Fminqb} thus corresponds to minimizing the expectation value of $\cT_D$ over all real unit vectors. As $\cT_D$ is real and diagonal, this minimum expectation value is exactly the smallest eigenvalue of $\cT_D$ (and hence of $\Nsym$), attained by the corresponding eigenvector normalized to unit length. Therefore, we see that the fidelity loss for a CPTP, unital qubit map $\Phi$ is given by
\begin{equation}
\eta_\Phi=1-\min_{|\psi\rangle\in\cC}F^2(|\psi\rangle,\Phi)=\frac{1}{2}(1-t_{\min}),
\end{equation}
where $t_{\min}$ is the smallest eigenvalue of $\Nsym$ corresponding to the map $\Phi$. Setting $\Phi=\RP\circ\cE\circ\cP$ gives $\eta_P$. Note that, for $\Phi$ with a Hermitian-closed set  \footnote{A set $\cK\equiv\{K_i\}$ is Hermitian-closed if $K_i\in\cK$ if and only if $K_i^\dagger\in\cK$.} of Kraus operators, as is the case for $\Phi\equiv \RP\circ\cE\circ\cP\sim\{PE_i^\dagger \cE(P)^{-1/2}E_j^\dagger P\}$, $\cT$ is symmetric so that $\Nsym=\cT$.

\section{Example: Amplitude damping channel}\label{sec:Example}

\begin{figure}[!htp]
	\includegraphics[width=0.52\textwidth]{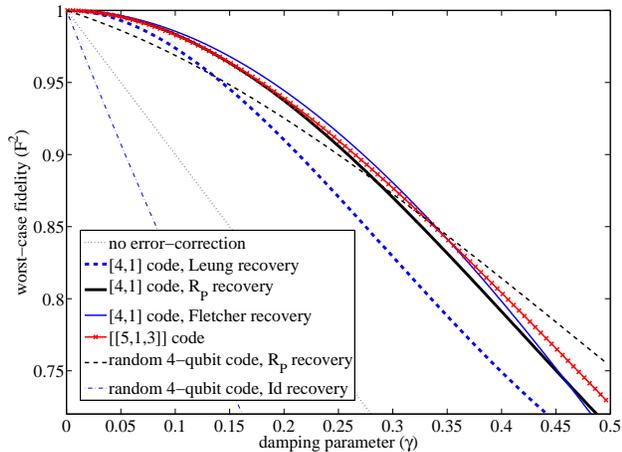}
	\caption{\label{fig:AmpDamp} Codes for the amplitude damping channel, for $0\leq \gamma\leq 0.5$.}
\end{figure}

As an example to illustrate our discussion so far, let us look at the noise channel considered in \cite{Leung}---the amplitude damping channel. The single-qubit amplitude damping channel $\cE_\text{AD}$ is the CPTP channel with Kraus operators
\begin{equation}
E_0=\left(
\begin{array}{cc}
1&0\\0&\sqrt{1-\gamma}
\end{array}
\right)\quad\text{and}\quad
E_1=\left(
\begin{array}{cc}
0&\sqrt\gamma\\0&0
\end{array}
\right),
\end{equation}
written in some qubit basis $\{|0\rangle,|1\rangle\}$. $\cE_\text{AD}$ can be thought of as describing energy dissipation for a system where $|0\rangle$ is the ground state, and $|1\rangle$ is some excited state. $\gamma$ is then the probability of a transition from the excited state to the ground state. In the absence of any encoding or recovery, the worst-case fidelity for a single qubit undergoing $\cE_\text{AD}$ falls off as $1-\gamma$, as $\gamma$ increases (see Fig. \ref{fig:AmpDamp}, line labeled ``no error correction").

A code that uses four physical qubits to protect a single qubit of information against amplitude damping noise was constructed by Leung et al. \cite{Leung}.
Assuming that the noise acts independently on the qubits, the four-qubit noise channel is just four copies of $\cE_\text{AD}$, i.e., $\cE_\text{AD}^{\otimes 4}$. The four-qubit subspace code constructed in \cite{Leung} is the span of the following two states:
\begin{align}
|0_L\rangle&\equiv\frac{1}{\sqrt 2}\left(|0000\rangle+|1111\rangle\right),\notag\\
\text{and}\quad|1_L\rangle&\equiv\frac{1}{\sqrt 2}\left(|0011\rangle+|1100\rangle\right).
\end{align}
$|0_L\rangle$ and $|1_L\rangle$ respectively represent the $|0\rangle$ and $|1\rangle$ states of the single qubit of information we want to encode in the four-qubit Hilbert space. We denote this code as the [4,1] code, where the first entry in the brackets corresponds to the number of qubits in the system, and the second entry is the number of qubits of information encoded in the system. It was shown in \cite{Leung} that this code satisfies the perfect QEC conditions for $\cE_{AD}^{\otimes 4}$, except for small corrections of order $\gamma^2$, and hence a recovery operation similar to $\Rperf$ can be constructed. We refer to this recovery map as the \emph{Leung recovery}. The worst-case fidelity for this code and recovery is plotted as a function of $\gamma$ in Fig. \ref{fig:AmpDamp}. Clearly, the [4,1] code is able to significantly raise the worst-case fidelity for the encoded qubit of information, as compared to the no error correction case. 

In the same figure, we also plot the worst-case fidelity using the transpose channel $\RP$ as the recovery operation, instead of the Leung recovery, for the same [4,1] code. We see that using the transpose channel as the recovery map gives a higher fidelity than the original Leung recovery.

For comparison, we also look at a recovery map for the [4,1] code constructed by Fletcher et al. in \cite{Fletcher07b}. This recovery map, which we refer to as the \emph{Fletcher recovery}, was originally optimized for an averaged measure of fidelity. We instead compute the worst-case fidelity for this recovery \footnote{The recovery map we used is given in Table I of \cite{Fletcher07b}. The Fletcher recovery map actually depends on two parameters $\alpha$ and $\beta$ which can be numerically optimized, for each value of $\gamma$, for the best recovery map. For simplicity, we set $\alpha=\beta=1/\sqrt{2}$ in our plot, which corresponds to the ``code-projected recovery" in \cite{Fletcher07b} with comparable performance as the fully optimized recovery.}, also plotted in Fig. \ref{fig:AmpDamp}. For small values of $\gamma$, the Fletcher recovery gives the best performance compared to the other recovery maps, despite being optimized for an averaged measure of fidelity. However, it only does marginally better than the transpose channel recovery.

We also compare the performance of the [4,1] code under these different recovery maps with that of a code that satisfies the perfect QEC conditions. 
The smallest code capable of perfectly correcting an arbitrary error on any single qubit, requires five qubits. The relevant noise channel now is $\cE_\text{AD}^{\otimes 5}$. The five-qubit code \cite{BDSW96,Laflamme96}, usually referred to as the [[5,1,3]] code \footnote{The first two entries in the double brackets mean the same as in the [4,1] code. The third entry is the distance parameter given by $2t+1$ where $t$ is the number of errors in the system the code can perfectly correct. The five-qubit code is capable of correcting an error on any qubit, so its distance parameter is equal to 3.}, satisfies the perfect QEC conditions for the CP channel comprising only the single-qubit (Pauli) errors in $\cE_\text{AD}^{\otimes 5}$. Using the corresponding $\Rperf$ as the recovery for the [[5,1,3]] code, we compute the worst-case fidelity for the noise channel $\cE_\text{AD}^{\otimes 5}$, for different values of $\gamma$. As the plot in Fig. \ref{fig:AmpDamp} shows, the [[5,1,3]] code performs better than the [4,1] code with Leung recovery, but the [4,1] code uses one qubit less to encode the same amount of information. The [4,1] code with the transpose channel as recovery has nearly identical worst-case fidelity as the [[5,1,3]] code, while the one with Fletcher recovery does slightly better for small values of $\gamma$. 

These observations clearly demonstrate the benefit of going beyond codes described by the perfect QEC conditions. Furthermore, while the [[5,1,3]] code is capable of perfectly correcting an arbitrary single-qubit error in a system subjected to \emph{any} noise channel, the comparison with the [4,1] code with various recovery maps clearly show the gain that one might achieve by adapting the codes and recovery to the noise channel in question.

Lastly, we also compute the worst-case fidelity for randomly generated four qubit codes, using the transpose channel as the recovery map. 
Computing $F^{2}$ for about 500 randomly selected codes took less than half an hour on a typical laptop computer. We plot the worst-case fidelity for the best code in Fig. \ref{fig:AmpDamp} (line marked ``random 4-qubit code, $\RP$ recovery"). 
For small values of $\gamma$, this random code does not do as well as the other codes discussed so far for the amplitude damping channel, but it still does significantly better than the case without error correction. Furthermore, for $\gamma\gtrsim 0.35$, our randomly generated code actually outperforms all the other codes. For comparison, we have also plotted the worst-case fidelity for this randomly generated code in the absence of the transpose channel recovery, i.e., with the identity channel as the recovery map (line marked ``random 4-qubit code, Id recovery"). One should keep in mind the ease with which the performance of the randomly generated code was achieved, due to the fact that the transpose channel is a near-optimal recovery map for \emph{any} code.

\begin{figure}[!htp]
	\includegraphics[width=0.52\textwidth]{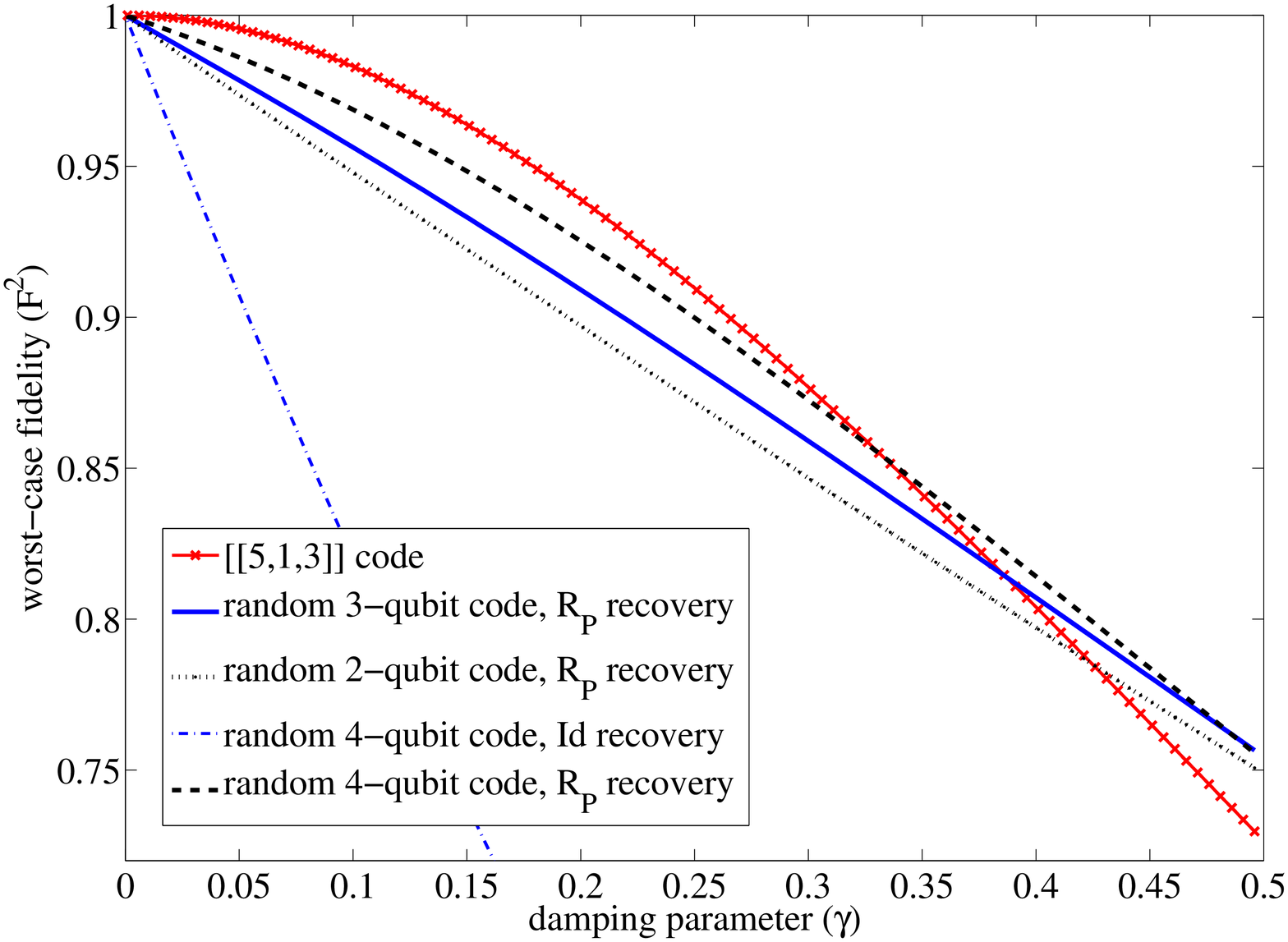}
	\caption{\label{fig:AmpDamp234} Randomly generated two-, three-, and four-qubit codes using the transpose channel as the recovery map. For comparison, we have also plotted the worst-case fidelity for the [[5,1,3]] code, and that of the randomly generated four-qubit code with no recovery (i.e., identity channel as recovery).}
\end{figure}

Finally, we also consider the possibility of constructing two-qubit and three-qubit codes for the amplitude damping channel. Because the transpose channel is near-optimal for any code, it can be used a good recovery map for the codes we generate, thus eliminating the need to search for a good recovery for every randomly selected code. The worst-case fidelity for the best codes we found are plotted in Fig. \ref{fig:AmpDamp234}. For comparison, we also plot the worst-case fidelities for the randomly generated four-qubit code mentioned in the previous paragraph, with the transpose channel and the identity channel as recovery maps. The corresponding graphs for the two- and three-qubit codes with identity channel as recovery are close to that of the four-qubit code. From the figure, we see that while the worst-case fidelity decreases as the number of physical qubits decreases, the two- and three-qubit codes in fact do not perform too badly compared to the four-qubit code or the [[5,1,3]] code. Such codes may be of relevance whenever the desire to lower resource requirements trumps the need for the best possible worst-case fidelity.

\section{Conclusions and open problems}\label{sec:Conc}

In this work, we demonstrated the crucial role the transpose channel plays in perfect QEC, and used it to formulate a simple approach to characterizing and finding AQEC codes. Compared to previous work based on numerically-generated recovery maps specific to the noise channel in question, the universal and analytically simple form of our transpose channel makes it particularly useful towards developing a better understanding of AQEC. While not being the optimal recovery in the case of AQEC codes, the near-optimality of the transpose channel provides a simple algorithm for identifying codes that satisfy some maximum fidelity loss requirements, without having to perform a difficult optimization over all recovery maps for every possible encoding. Furthermore, our approach, founded upon the worst-case fidelity rather than an averaged measure of fidelity, provides the often desirable guarantee that the code found is able to protect \emph{all} information that can be stored in the code with some minimum fidelity. We have also shown that the case of qubit codes is particularly easy to handle, and our method of computing the worst-case fidelity for a CPTP qubit map might be useful in contexts beyond our present discussion.

There are many interesting related open problems. An immediate question is whether the gap present in our AQEC conditions between the necessary and sufficient conditions (arising from the inequality in Corollary \ref{cor:iff}) can be reduced, either by improving the bound in Theorem \ref{thm:Fid}, or by using a different recovery map that might perform better than the transpose channel. It would be very interesting if a simple and universal recovery map could be found, for which the dimension of the code does not appear in the worst-case fidelity. There is also the question of whether it might be possible to extend our efficient method of computing the worst-case fidelity for qubit codes to higher dimensional codes and more general channels. Finally, we expect that the transpose channel can also be used to study approximate codes more general than subspace codes, like for example, OQEC codes \cite{OQECC06}.

Another important problem is to figure out whether the transpose channel can be easily implemented using measurements and gates. In the case of perfect QEC, the transpose channel (or equivalently $\Rperf$) can be implemented simply using syndrome measurements and conditional gates (see for example, \cite{NCBook}). In order for AQEC codes to be useful for computational or communication tasks, it must be possible to implement the recovery operation using physical operations that are not overly complicated or demanding in resources. This is in fact another advantage of our analytical approach over numerically-constructed recovery maps for which no practical implementation structure may be apparent (although, see \cite{Fletcher07a}).

AQEC provides a new and mostly unexplored arena of possibilities for the design of codes to protect information from noise for use in quantum information processing tasks. Our work provides an analytical characterization of AQEC and further analytical understanding will undoubtedly prove invaluable towards unlocking the full potential of AQEC.

\acknowledgments
We would like to thank David Poulin for introducing us to the problem of approximation quantum error correction, and for many insightful discussions. This work is supported by NSF under Grant No. PHY-0803371.

\bibliographystyle{apsrev}

\end{document}